\documentclass[11pt]{article}
\usepackage{amsmath,amsfonts,amssymb,amsthm,bm,stackrel,graphicx}
\usepackage{natbib}

\newcommand{\kth}[1]{${#1}^{\textrm{th}}$}

\begin{document}
\title{Recovering Direct Effects in Genetics: A Comparison} 
\author{MATTHEW SPERRIN$^\ast$\\[4.1pt]
{\it \small Department of Mathematics and Statistics, Lancaster University, UK}
\\[2pt]
{\small Tel: +44 (0)1524 594746}\\[2pt]
{\small m.sperrin@lancaster.ac.uk}\\[7pt]
 THOMAS JAKI\\[4.1pt]
{\it \small Department of Mathematics and Statistics, Lancaster University, UK}
\\[2pt]
{\small Tel: +44 (0)1524 592318}\\[2pt]
{\small jaki.thomas@gmail.com}\\
}
\maketitle
\renewcommand{\thefootnote}{\fnsymbol{footnote}}
\footnotetext{$^\ast$ To whom correspondence should be addressed.}

\begin{abstract}
In genetics it is often of interest to discover single nucleotide polymorphisms (SNPs) that are directly related to a disease, rather than just being associated with it.  Few methods exist, however, addressing this so-called `true sparsity recovery' issue.
In a thorough simulation study, we show that for moderate or low correlation between predictors, lasso-based methods perform well at true sparsity recovery, despite not being specifically designed for this purpose.  For large correlations, however, more specialised methods are needed. Stability selection and direct effect testing perform well in all situations, including when the correlation is large.
\end{abstract}

\section{Introduction}
In genetic studies where the density of measured single nucleotide polymorphisms (SNPs) is high, it is of interest to recover SNPs that are directly affecting a response.  A SNP with a direct effect is defined as one that has an effect on the response not caused by its correlation with any other measured SNP.  Unfortunately, there can be many SNPs associated with the response due to high linkage disequilibrium (LD) or correlation with a SNP with a direct effect.  Define the `true sparsity pattern' to be the set of SNPs (or more generally predictors) that have a direct effect on the response.  This paper compares the ability of various methods to recover the true sparsity pattern.

Consider a study with $n$ participants typed at $p$ SNPs. 
The SNPs and the response are assumed to be related through
\begin{equation} \label{basic-reg}
g(E[\bm{Y}]) = \bm{\eta} = \bm{X\beta} + \bm{\delta},
\end{equation}
where $\bm{\eta}$ is an $n \times 1$ response vector explained by an $n \times p$ matrix $\bm{X}$ (the SNPs) through an unknown $p \times 1$ coefficient vector $\bm{\beta}$ with $n \times 1$ noise vector $\bm{\delta}$. The response of interest is $\bm{Y}$ and $g(\cdot)$ is the link function \citep[see][]{mccullagh89}.  We are interested in a sparse solution where only some $\beta_j \neq 0$. For a review of sparsity methods  see \cite{fan10}.  SNPs with direct effects are difficult to find when the predictors are correlated or $p > n$. In these situations there may exist $\bm{\beta}'$ and $\bm{\beta}''$ such that $\bm{X\beta}' = \bm{X\beta}''$ but $\bm{\beta}' \neq \bm{\beta}''$, causing an identifiability problem \citep[e.g.][]{candes07}.
It can therefore be difficult to identify which of a highly correlated group of predictors possesses a direct effect.

We consider the case where the predictors and response are binary.  SNPs can be expressed in binary format despite having three levels by coding the dominant effect and recessive effect in two separate predictors. This allows us to capture dominant, recessive and additive effects where the latter is represented as significance in both the dominant and recessive effect predictors. In genetic studies, the responses are also frequently binary as, for example, in case control studies. 

In this paper we compare the lasso, screen and clean, stability selection, direct effect testing, and the elastic net; these methods are briefly described in Section \ref{sec:methods}. Of interest is their ability to recover true sparsity patterns in simulations. The methods are compared against Fisher's exact test, the standard test for binary association.  Section \ref{sec:design}  describes the simulations and presents the results. A comparison on  real data is carried out in Section \ref{sec:data}; we conclude in Section \ref{sec:discuss}.

\section{Sparsity Methods} \label{sec:methods}
\subsection{Lasso}
The lasso is a regression technique that performs variable selection \citep{tibshirani96}. Estimation is done by minimising a penalised residual sum of squares function,
\begin{equation} \label{eq:lasa}
\hat{\bm{\beta}} =  \ \stackrel[\bm{\beta}]{}{\textrm{argmin}} (\bm{Y} - \bm{X\beta})^2 + \lambda \sum_{j=1}^p|\beta_j| ,
\end{equation}
where $\lambda$ is a tuning parameter.  If $\lambda = 0$, the lasso is identical to ordinary least squares while a larger value of $\lambda$ will cause some coefficients to be zero, which corresponds to selection, and the remaining coefficients to be shrunk towards zero.  The tuning parameter $\lambda$ can be selected through cross validation or standard model selection criteria.

While lasso regression can be computed efficiently using the least angle regression algorithm (LARS) of \cite{efron04} it suffers from difficulty assigning significance to predictors.  \cite{meinshausen08a} describes this notion as ill-posed; it may be obvious that an effect is present, but  significance is  lost in trying to assign the effect to a specific predictor.  This leads to the naive inclusion of all those predictors with nonzero coefficients in the final sparse model.

\subsection{Screen and Clean}
Screen and clean \citep{wasserman08} is an extension to the lasso that uses a two stage approach to overcome two issues with the lasso: first, the difficulty of significance testing in the presence of a large number of potentially correlated predictors, and second, the tendency of the lasso to underestimate effect sizes \citep[e.g.][]{radchenko09}.
The first `screen' stage fits a lasso regression to half of the data,  producing a collection of  predictors.  The second `clean' stage fits ordinary least squares regression to the other half of the data, using the predictors carried forward from the first stage. This reduced set of coefficients makes significance testing feasible, and avoids effect sizes being underestimated. Significance testing is carried out with a Bonferroni correction on the reduced set.

\subsection{Stability Selection}
Stability selection \citep{meinshausen08b} tackles the problem of significance testing for predictors by resampling. Each resampled dataset is generated by selecting half of the observations at random (without replacement). For each subsample lasso regression is fit, and the predictors with nonzero coefficients are recorded as those selected.  Let $m_j$ be the count of times the \kth{j} predictor is nonzero.   For each predictor $\bm{X}_j$, the proportion of times that predictor is present in the fitted model is given by $\pi_j = m_j/B$.  Predictor $\bm{X}_j$ is included in the final model if $\pi_j > \pi_{\textrm{thr}}$, where $\pi_{\textrm{thr}}$ is a tuning parameter.  The set of predictors recovered is insensitive to the choice of $\pi_{\textrm{thr}}$, for $\pi_{\textrm{thr}} \in (0.6,0.9)$ \citep{meinshausen08b}. Note that other subset selection method then the lasse can be considered; we use lasso regression for comparability with the other methods.

\subsection{Elastic Net}
The elastic net \citep{zou05} combines lasso and ridge regression \citep{hoerl88}. 
Like the lasso, it minimises a penalised residual sum of squares function, but there is an additional quadratic penalty term:

\begin{equation} \label{eq:elastic}
\hat{\bm{\beta}} =  \ \stackrel[\bm{\beta}]{}{\textrm{argmin}} (\bm{Y} - \bm{X\beta})^2 + \lambda \alpha \sum_{j=1}^p|\beta_j| + \lambda (1-\alpha) \sum_{j=1}^p\beta_j^2.
\end{equation}
As a consequence, amongst a set of highly correlated predictors, the elastic net tends to include either all or none of those predictors in the model. This is in contrast to the lasso, which tends to select a single predictor from a group of highly correlated predictors that are correlated with the response. 

\subsection{Direct Effect Testing}
Direct effect testing (DET) \citep{sperrin09} detects direct effects, and calculates the probabilities that each predictor is the true origin of each direct effect. 
The first stage of DET identifies direct effects by using lasso regression to separate direct effects from indirect effects. To calculate the significance of an effect, each direct effect is attributed, automatically by the lasso, to a specific predictor. 
The second stage of DET then incorporates the uncertainty in the predictor that is the true origin of each direct effect, by giving, for each predictor $\bm{X}_k$ on which an effect is observed,  
$$
\Pr[\textrm{$\bm{X}_j$ true effect}|\textrm{Effect observed in stage one on $\bm{X}_k$}].
$$
for a collection of predictors $\bm{X}_j$ in high correlation with $\bm{X}_k$.

DET therefore provides a method to carry out fine mapping by distinguishing between direct and indirect effects, and quantifying the uncertainty associated with this distinction.

\subsection{Pre-Screening Methods}
 In our simulations, we  restrict the number of predictors to numbers in the thousands. Genetic data, however, frequently involve several hundred thousand predictors.   \cite{fan08} suggest an initial step in which the dimensionality is reduced to a more manageable level using ``sure independence screening''. This is followed by application of the other methods described in this section. In our investigations we will assume that, when appropriate, some dimension reduction approach has been applied first to reduce the number of predictors.

\section{Design and Results of Simulations} \label{sec:design}

A range of  simulations are carried out to study the properties of the methods introduced above.  Situations involving strong and weak correlations, including both serially correlated and clustered data, are considered and consistency properties of the methods on these simulated data are investigated.    We begin by explaining how the various methods are applied, how significance is determined, and introduce some of the measures used to compare the methods.

For the lasso, the penalty is selected using cross validation;  a  find is  recorded when a predictor receives a nonzero coefficient. For screen and clean,  the lasso penalty is selected by cross validation and significance is determined using a Bonferroni correction. The results for stability selection are based on 100 resamples with the lasso penalty determined by cross validation and the significance threshold set at $\pi_{\textrm{thr}} = 0.75$. We calculate elastic net solutions for three values of $\alpha$: 0.9, 0.75 and 0.5. The penalty term $\lambda$ is chosen via cross validation. Only $\alpha=0.5$ is displayed in the proceeding simulations. As $\alpha \rightarrow 1$ the elastic net solution becomes identical to the lasso, hence elastic net solutions with larger $\alpha$ fall between the elastic net solutions displayed, and the lasso solutions. For DET, significance of each predictor is calculated using a Bonferroni correction.  For stage two, let $p_{\textrm{max}}$ be the probability of the best contender for the origin of a given direct effect, then a `find' is defined as any other predictor that has a probability of origin that is at least $10\%$ as large as the best contender, i.e.\ $0.1p_{\textrm{max}}$.  We also include results for Fisher's exact test, with a Bonferroni correction, as a baseline comparison.

The comparisons are not designed to be an indicator of which methods are universally best.  Each method is designed to deal with different scenarios, and here only the ability of the methods to recover true sparsity patterns is considered.  Moreover, it is difficult to make the comparisons fair, for example DET has multiple chances to make a find per significant effect identified, while formal significance testing is not used in declaring significance for the lasso.

 The response for each individual observation, $Y_i$, is generated according to $\Pr[Y_i = 1] = \mu_i$, where
$$
\log \left( \frac{\mu_i}{1- \mu_i} \right) = \alpha + \beta_{j_1} X_{ij_1} + \ldots +   \beta_{j_K} X_{ij_K},
$$
so that the response is related to a subset of $K$ predictors with indices $\{j_1,\ldots,j_K\}$. The intercept $\alpha$ is chosen 
to provide approximately equal numbers of observations with $Y_i = 0$ and $Y_i = 1$; in this way we intend to replicate a case-control study.
The $\beta_{j_k}$'s are chosen to represent large effect sizes so that the differences between methods can be most easily seen; we use $\beta_{jk} = 0.81$ and $\beta_{jk} = 0.405$. These correspond to an absolute risk increase of 20\% and 10\% respectively, in the case control framework.
 The influential predictor(s) are chosen at random on each simulation.

We have generated data in a reasonably simple fashion: for example by considering first-order auto-regressive predictors.  More realistic correlation patterns for genetic data can be generated using real LD patterns. However, this does not give us the flexibility to change the extent of the correlation that we get from the methods used here. Simulations based around real datasets which therefore contained more realistic LD patterns where also undertaken. As the results were qualitatively the same as those under the simpler situations they are not presented.

If a method finds a predictor that was truly used to generate the response,  this is called a `true find'.  Any other predictors found are declared `false finds'.   A `strong false find' is recorded whenever at least one incorrect predictor is included in the model; a `strong true find' means all the causal predictors are recovered.  A `perfect find' means the model includes all causal predictors and no other predictors (i.e.\ occurrence of a strong true find and non-occurrence of a strong false find).  A standard definition of false discovery rate (FDR),
$
\textrm{FDR} = \frac{\textrm{Number of false finds}}{\textrm{Total number of finds}},
$
is also used. These definitions are summarised in Table \ref{tab:finddef}. 

Since in genetics it is often of interest to detect predictors in high LD with a causal variant, we also carry out simulations in which the definition of a `true find' is relaxed so that a true predictor is declared to have been found provided any predictor with a correlation of at least 0.9 to the true predictor has been found. We denote the results corresponding to this definition the HCT (highly correlated with the true predictor) cases. These results are only included for some cases, as the omitted scenarios were qualitatively the same. To enhance readibility, in all proceeding Figures, false finds are displayed on a log scale, whilst the remaining measures are displayed on a linear scale.

\begin{table}
\caption{Summary of find definitions}
\begin{center}
\begin{scriptsize}
\begin{tabular}{r|l}
Name &Description\\
\hline
True find &True predictor found\\
Strong true find &All true predictors found\\
False find &False predictor found\\
Strong false find &At least one false predictor found\\
Perfect find &Strong true find and no strong false find\\
FDR &False discovery rate (false finds / total finds)
\end{tabular}
\label{tab:finddef}
\end{scriptsize}
\end{center}
\end{table}
The `glmnet' function in the `glmnet' library \citep{glmnet}  in R \citep{R} was used to calculate lasso and elastic net solutions.  All other code was written by the authors.


\subsection{Serially Correlated Data} \label{sec:serial}


To generate serially correlated data, denoting the correlation by $\rho$, the approach used is:
\begin{enumerate}
\item Generate random binary realisations for $X_{i,1}$, $i=1,\ldots,n$, according to a Bernoulli(0.5) distribution,  where $n$, the sample size, is set to $n=1000$ here.
\item For $j=2,\ldots,p$, where $p$ is the number of predictors, set to $p=400$ here, generate binary realisations for $X_{i,j}$, $i=1,\ldots,n$, via
$$
X_{i,j} = \left\{ \begin{array}{cc}
X_{i,j-1} &\textrm{with prob. $\rho$},\\
0 &\textrm{with prob. $(1-\rho)/2$},\\
1 &\textrm{with prob. $(1-\rho)/2$}.\\
\end{array}\right.
$$
\item Randomly assign a causal predictor with coefficient 0.81, repeating this ten times on each dataset.
\item Repeat steps 1--3 100 times to obtain 100 different datasets, and hence 1000 simulations in total. 
\end{enumerate}
This procedure is repeated for a range of serial correlations, from $\rho = 0$ to $\rho = 0.99$.

\begin{figure}[!htbp]
\begin{center}
\includegraphics[width=14cm]{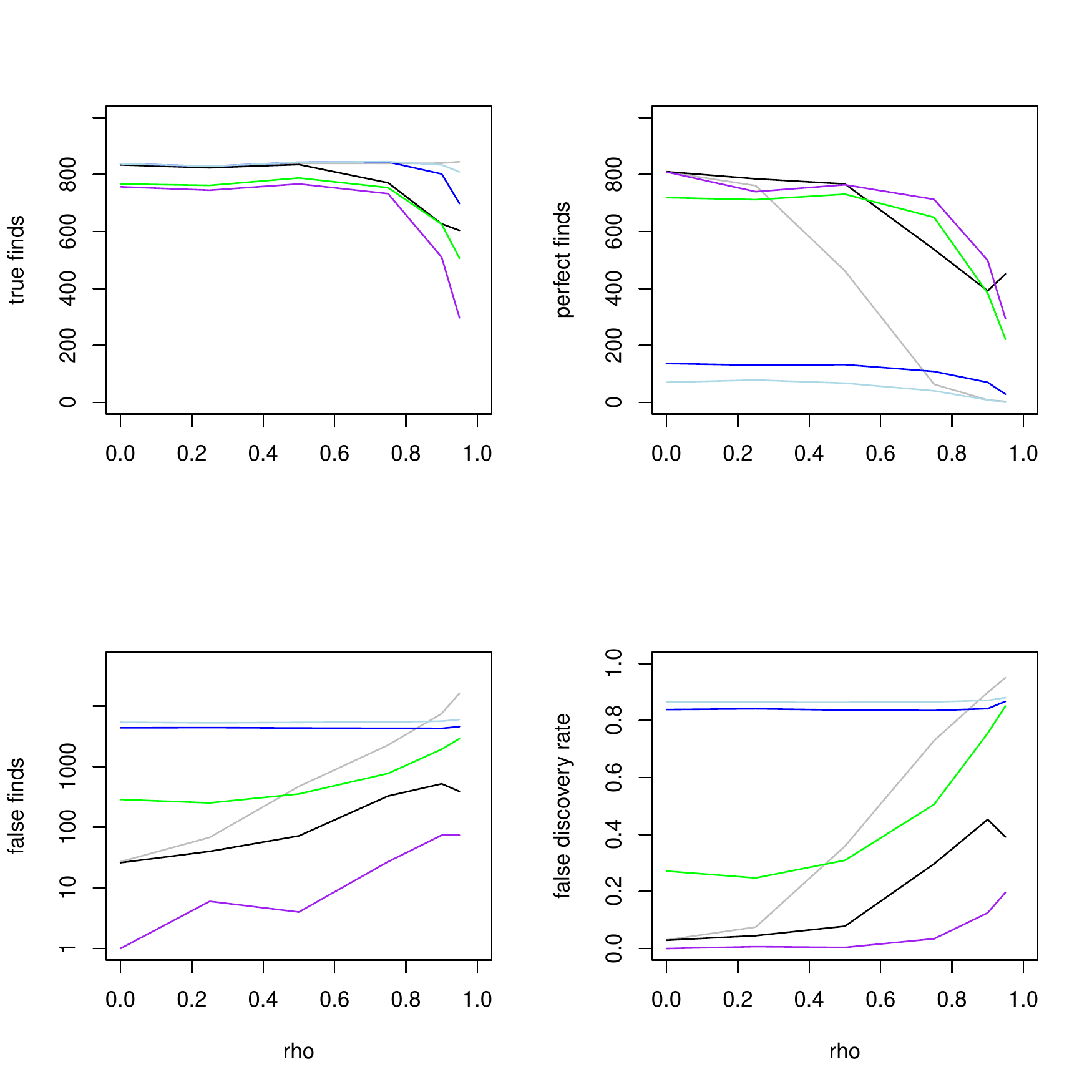}
\caption{Serial correlation: Comparing true finds (top left), perfect finds (top right), false finds (bottom left) and false discovery rate (bottom right) for different values of $\rho$. Grey line: Fisher test, black line: DET, blue line: lasso, light blue line: elastic net, green line: screen and clean, purple line: stability selection.}
\label{serial-g1}
\end{center}
\end{figure}

Figure \ref{serial-g1}  illustrates the simulated true finds, perfect finds, false finds and FDRs for serial correlations ranging from 0 to 1. 
Most of the methods deteriorate in fairly similar ways as the correlation approaches one.  The exception is DET, whose perfect find rate and FDR do not become worse as the correlation becomes very large.  Indeed, once the correlation becomes close to one, DET has the highest number of perfect finds. This is a consequence of the second stage of DET accounting for the high correlation by giving a large set of potential predictors for each detected effect.  It is only once $\rho > 0.75$, however, that including stage two of DET gives additional benefits over stage one alone (DET stage one not shown). For small to moderate correlation, stability selection gives the highest perfect find rate and lowest FDR. Figure \ref{serial-g1} also demonstrates poor false find control of the elastic net and Fisher test, although the Fisher test, unsurprisingly, copes well at low correlations.

\begin{figure}[!htbp]
\begin{center}
\includegraphics[width=14cm]{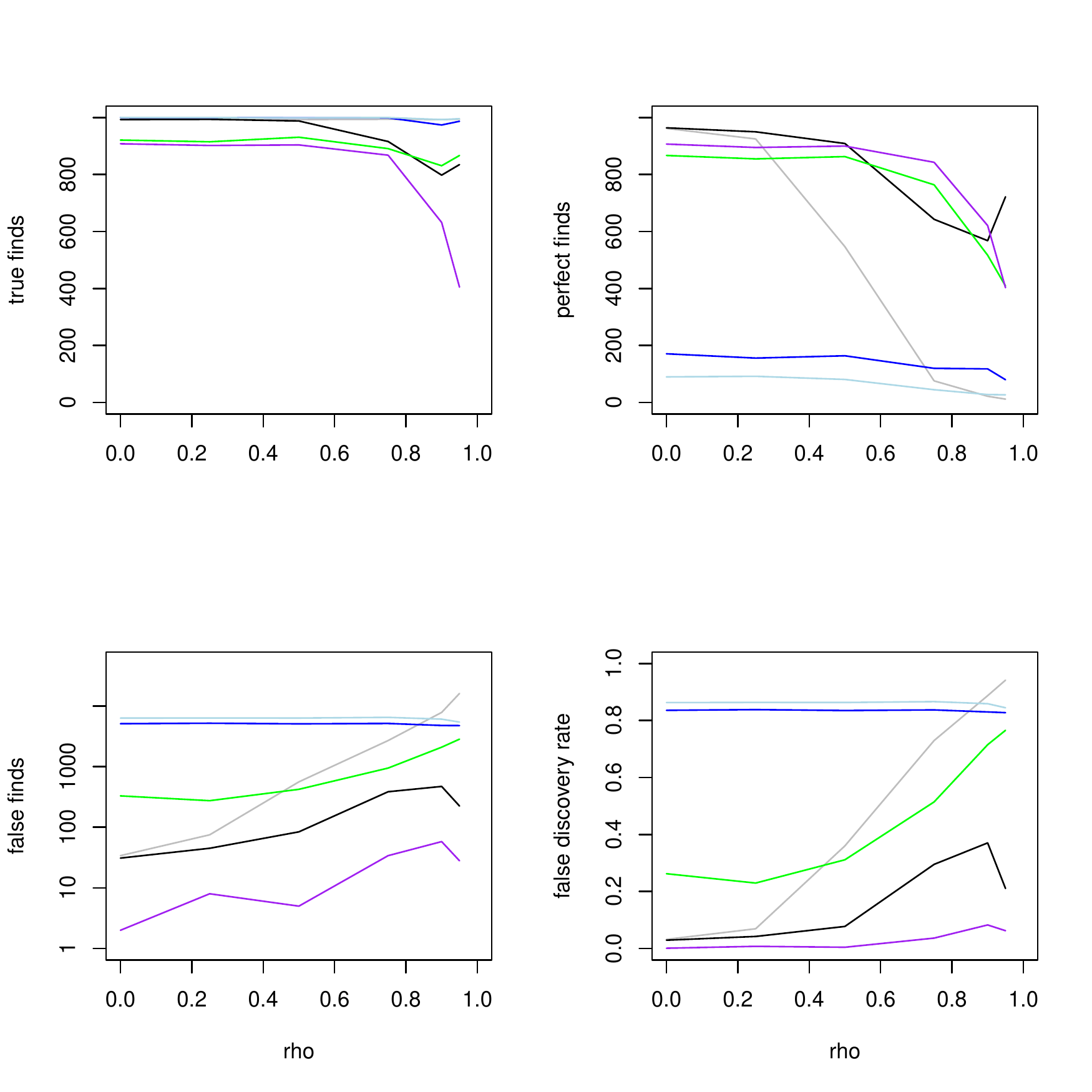}
\caption{Serial correlation (HCT case): Comparing true finds (top left), perfect finds (top right), false finds (bottom left) and false discovery rate (bottom right) for different values of $\rho$. Grey line: Fisher test, black line: DET, blue line: lasso, light blue line: elastic net, green line: screen and clean, purple line: stability selection.}
\label{serialsoft-g1}
\end{center}
\end{figure}

Figure \ref{serialsoft-g1}  illustrates the HCT case. Stability selection benefits the most from this relaxation of the definition of a find: the deterioration in its accuracy for large correlation is prevented. Lasso, the elastic net and the Fisher test do not appear to benefit. 

\begin{figure}[!htbp]
\begin{center}
\includegraphics[width=14cm]{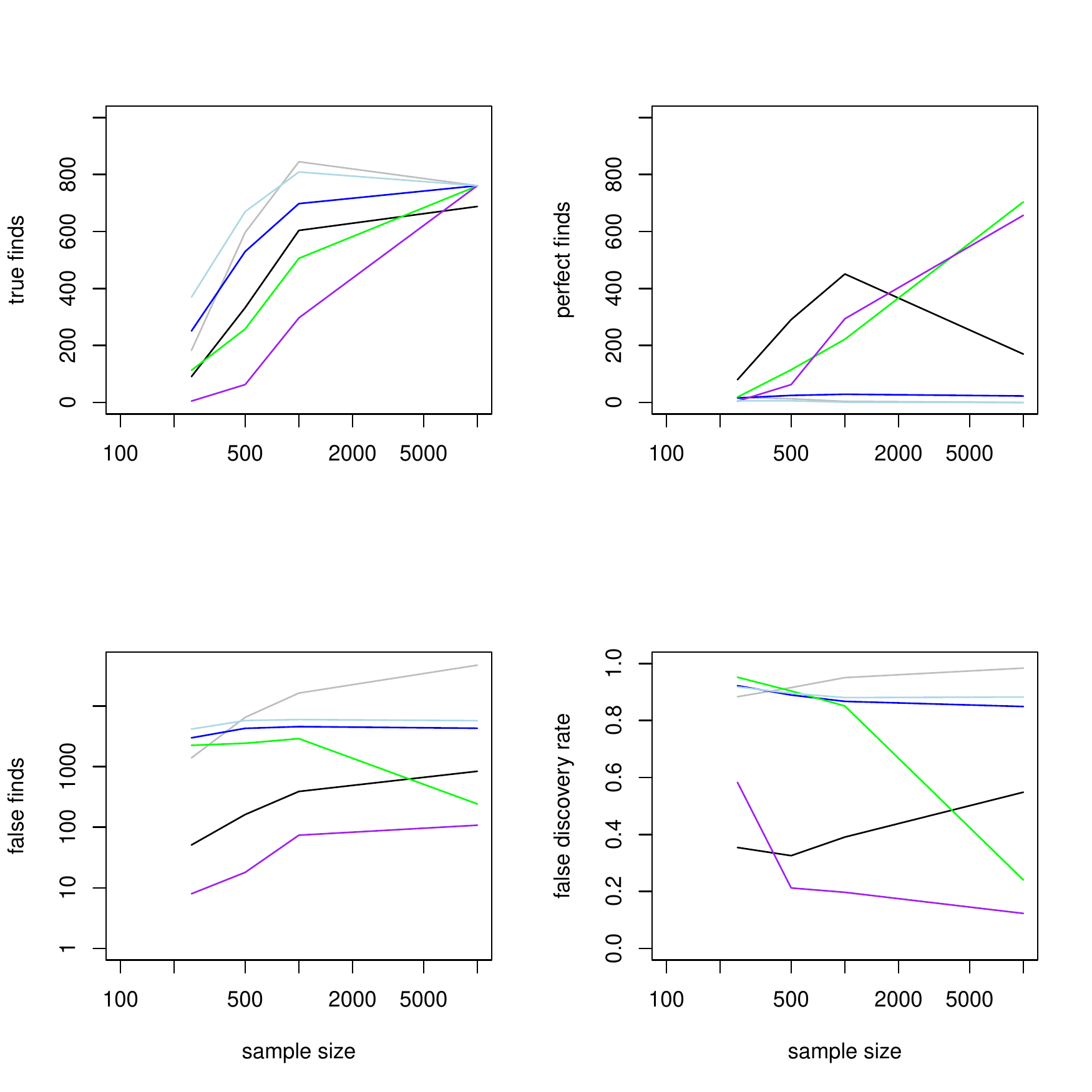}
\caption{Comparing true finds (top left), perfect finds (top right), false finds (bottom left) and false discovery rate (bottom right)  for differing sample sizes $n$, on the log scale. Serial correlation $\rho=0.9$. Grey line: Fisher test, black line: DET, blue line: lasso, light blue line: elastic net, green line: screen and clean, purple line: stability selection.}
\label{cons-95-g1}
\end{center}
\end{figure}
\begin{figure}[!htbp]
\begin{center}
\includegraphics[width=14cm]{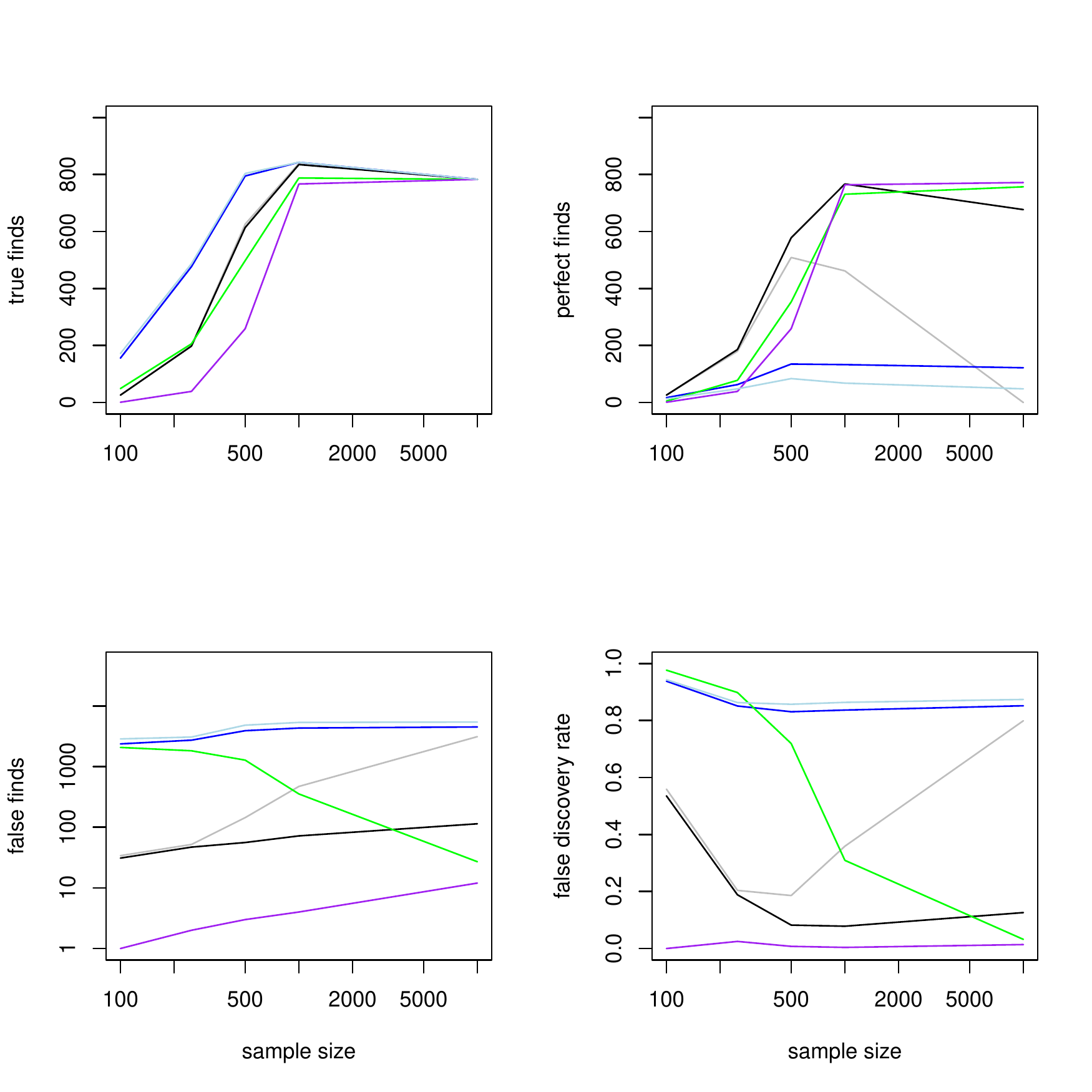}
\caption{Comparing true finds (top left), perfect finds (top right), false finds (bottom left) and false discovery rate (bottom right)  for differing sample sizes $n$, on the log scale. Serial correlation $\rho=0.5$. Grey line: Fisher test, black line: DET, blue line: lasso, light blue line: elastic net, green line: screen and clean, purple line: stability selection.}
\label{cons-50-g1}
\end{center}
\end{figure}

To study consistency, the same data generation as above was used and the correlation fixed while the sample size, $n$, varied. Figure \ref{cons-95-g1} gives true finds, perfect finds, false finds and FDRs for the methods with high correlation, $\rho=0.95$. Figure \ref{cons-50-g1}  gives the same for correlation $\rho=0.5$. In both cases the number of predictors $p=400$.

Screen and clean has the best consistency properties in the high correlation scenario, since the perfect find rate increases, and the false discovery rate decreases as the sample size increases.  Stability selection also performs well in the high correlation scenario since the perfect find rate increases, and the false discovery rate is controlled (although it does not appear to decrease as fast when $n$ increases).  DET performs poorly as sample size increases in the large correlation case; a large number of false finds are generated for large $n$, and the perfect find rate does not increase.  For lasso, elastic net and the Fisher test, there appears to be no gain in having a sample size larger than 1000.

For the low correlation case (Figure \ref{cons-50-g1}), all the methods considered have good consistency properties.  DET outperforms the other methods in terms of perfect finds and FDR for small $n$; stability selection and screen and clean are the better performers on these measures for larger sample sizes $n$. 
\subsection{Clustered Data}


Clustered data is generated with $p=400$ predictors, divided into clusters of size $k$, for a range of cluster sizes from $k=1$ to $k=10$. When necessary extra predictors are added to the last cluster to obtain the desired total number of predictors. Within cluster correlation is set to $\rho$ and while different clusters are independent. The data generation procedure is as follows:
\begin{enumerate}
\item Generate random binary realisations for $X_{i,j}$, $i=1,\ldots,n$, where the sample size is set to $n=1000$, according to a Bernoulli(0.5) distribution for the first predictor $\bm{X}_j$ in each cluster. Hence this step is carried out $p/k$ times.
\item For each subsequent $\bm{X}_j$ in each cluster, generate binary realisations for $X_{i,j}$, $i=1,\ldots,n$ via
$$
X_{i,j} = \left\{ \begin{array}{cc}
X_{i,j-1} &\textrm{with prob. $\rho$},\\
0 &\textrm{with prob. $(1-\rho)/2$},\\
1 &\textrm{with prob. $(1-\rho)/2$}.\\
\end{array}\right.
$$
Hence this is carried out $k-1$ times in each cluster.
\item Randomly assign a causal predictor with coefficient 0.81, repeating this ten times on each dataset.
\item Repeat steps 1--3 100 times to obtain 100 different datasets, and hence 1000 simulations in total. 
\end{enumerate}

\begin{figure}[!htbp]
\begin{center}
\includegraphics[width=14cm]{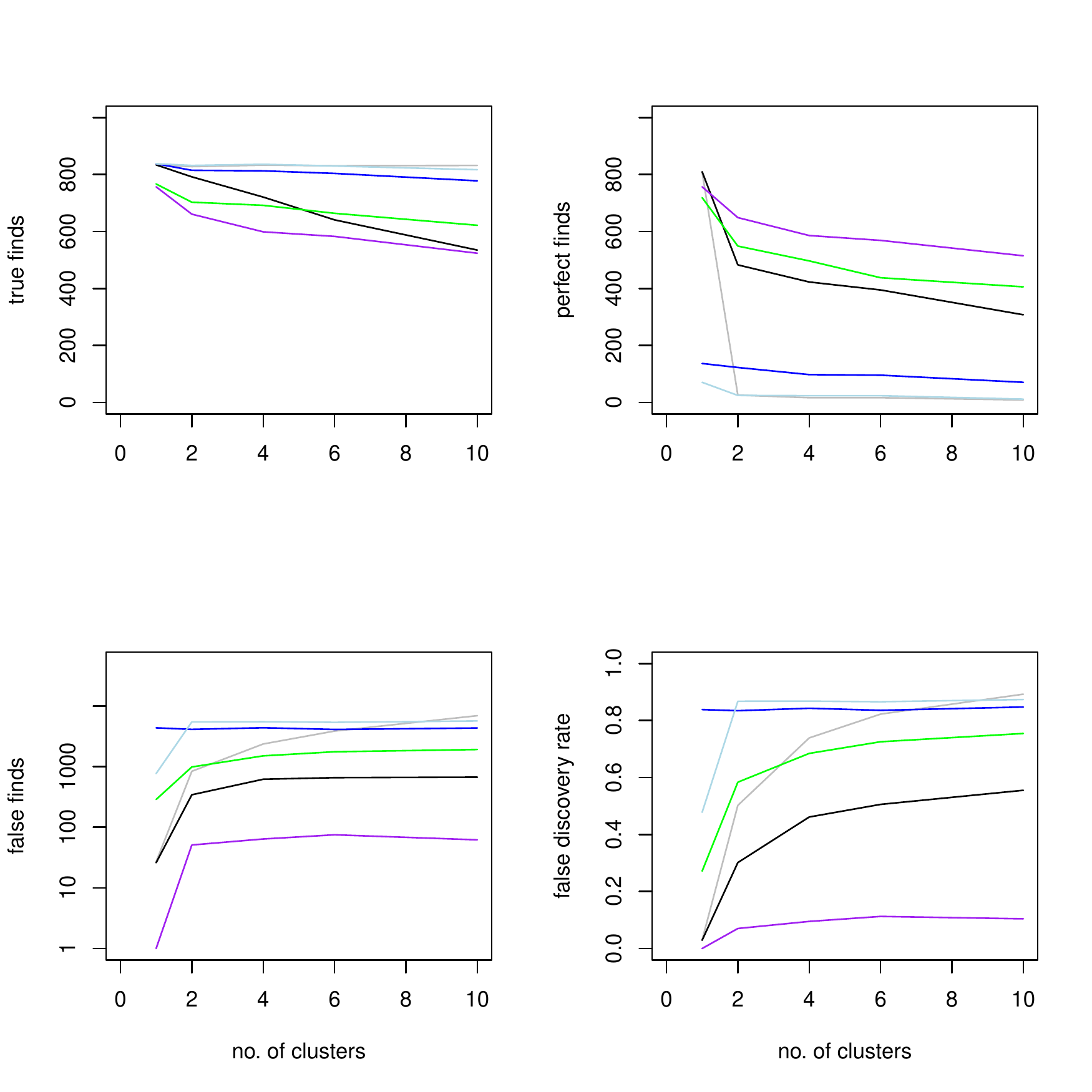}
\caption{Comparing true finds (top left), perfect finds (top right), false finds (bottom left) and false discovery rate (bottom right)  for different values of cluster sizes, within cluster correlation 0.9. Grey line: Fisher test, black line: DET, blue line: lasso, light blue line: elastic net, green line: screen and clean, purple line: stability selection.}
\label{clus09-g1}
\end{center}
\end{figure}

Figure \ref{clus09-g1} visualises the results of various cluster sizes, for a within cluster correlation $\rho = 0.9$.  Stability selection controls the FDR and maintains a large number of perfect finds, out-performing all the other methods. Cluster size has little effect on the performance of lasso and elastic net, but a large detrimental effect on the performance of the Fisher test.

\begin{figure}[!htbp]
\begin{center}
\includegraphics[width=14cm]{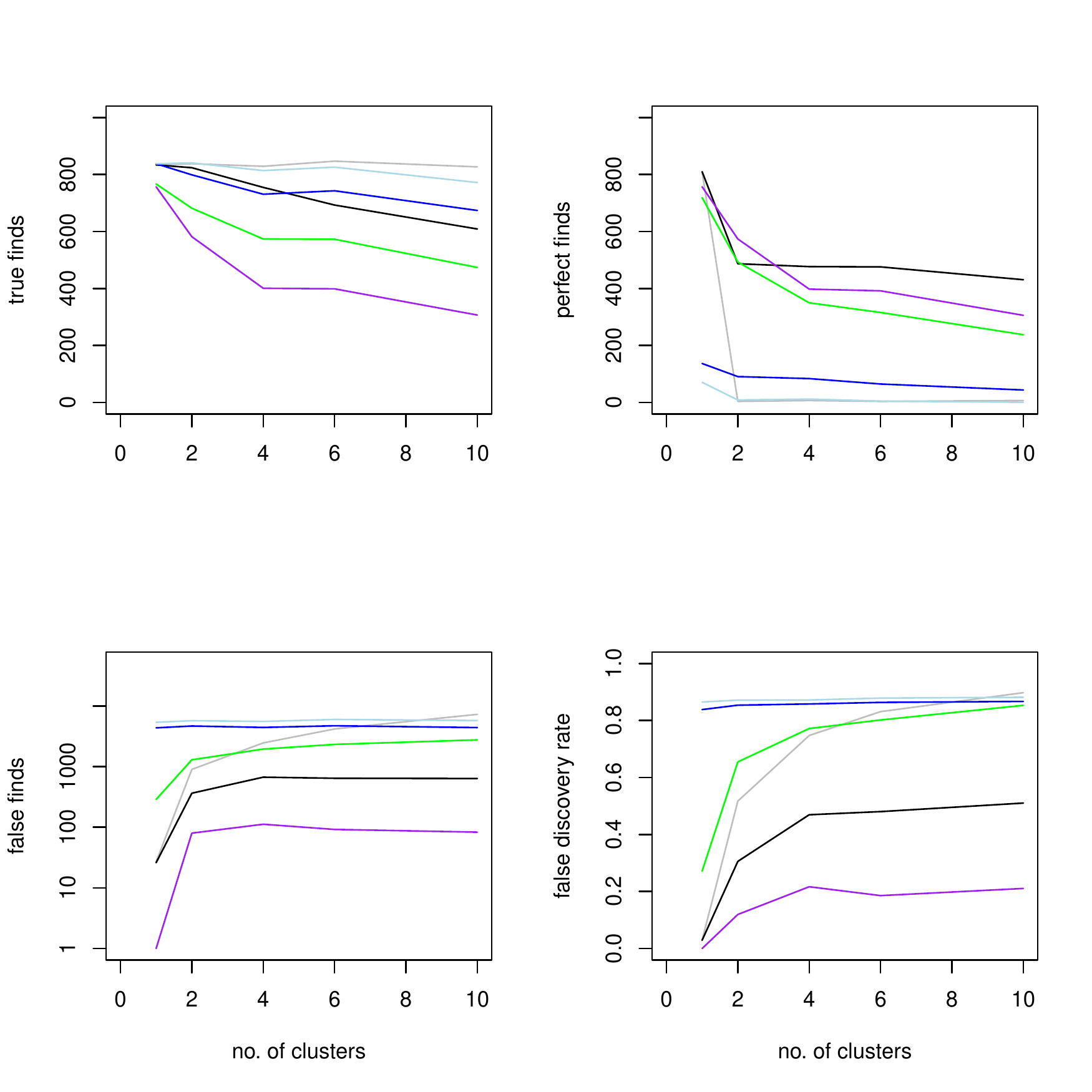}
\caption{Comparing true finds (top left), perfect finds (top right), false finds (bottom left) and false discovery rate (bottom right)  for different values of cluster sizes, within cluster correlation 0.95. Grey line: Fisher test, black line: DET, blue line: lasso, light blue line: elastic net, green line: screen and clean, purple line: stability selection.}
\label{clus095-g1}
\end{center}
\end{figure}

Figures \ref{clus095-g1} repeats the cluster size study but within cluster correlation is set to $\rho = 0.95$.  The larger correlation causes stability selection to struggle to make perfect finds, but it does continue to control the FDR.  DET now achieves the largest number of perfect finds, and the second best FDR control. Comparing Figures \ref{clus09-g1} and \ref{clus095-g1} shows that increasing the correlation has little effect on the performance of the lasso, elastic net or Fisher test.

Consistency of the various methods was also considered in the clustering framework.  The results are qualitatively similar to those in the serial correlation framework, hence are omitted.


\section{Real Data Example} \label{sec:data}
We compare the  methods on a dataset arising from a case-control study investigating whether there are SNPs that act as biomarkers for hypersensitivity (yes or no) to abacavir, a treatment for HIV \citep{kelly06}. We will use the combined data from two global trials \citep{hetherington02} in our investigation. 

The initial dataset consisted of 856 SNPs that have been arbitrarily numbered here and are spread across the 22 autosomal chromosomes. To obtain binary data, we separate the dominant and recessive effect of each SNP into two predictors and then remove any binary predictor whose prevalence of 0s or 1s is below 5\% giving $p=1373$ binary predictors for our evaluations. The methods are applied and the SNPs that have been declared `finds' are recorded in Table \ref{tab:abacavir}. `Perfect finds' cannot be considered here as the true relationship is unknown.

\begin{table}
\caption{SNPs in the abacavir data that were declared true finds by the various methods considered.}
\begin{center}
\begin{tiny}
\begin{tabular}{c|c|c|c|c|c|c|c|c|c}
    &    &      &Stab     &  &Elastic &\\       
SNP &DET &Lasso &Sel & S \& C  &net &Fisher\\
\hline
98D  & & & &x& & \\\hline
133D & & & &x& & \\\hline
150D & & & & & &x\\\hline
229D &x&x&x&x&x&x\\\hline
229R & & & & & &x\\\hline
235D & &x& & & &x\\\hline
331R & & & & & &x\\\hline
334D & &x& & &x&x\\\hline
335R & & & &x&x&x\\\hline
336D &x&x&x& &x&x\\\hline
340D & & & & & &x\\\hline
344R & & & &x& & \\\hline
349D & & & & & &x\\\hline
391R & &x& & & &x\\\hline
392R & &x& & & &x\\\hline
406D & & & & & &x\\\hline
440D & & & & & &x\\\hline
440R & & & & & &x\\\hline
443D & & & &x&x&x\\\hline
443R & &x& &x&x&x\\\hline
488R & & & &x& & \\\hline
489D & & & & &x&x\\\hline
489R & & & & & &x\\\hline
529D & &x& & & & \\\hline
613D & & & & & &x\\\hline
642D & & & &x& & \\\hline
738D & &x& & &x&x\\\hline
757D & & & & & &x\\\hline
758D & &x& & &x&x\\\hline
811D & & & & & &x\\\hline
\end{tabular}
\label{tab:abacavir}
\end{tiny}
\end{center}
\end{table}

The most conservative methods are DET and stability selection, both of which find the same two SNPs significant --- the dominant effects on SNP 229 and SNP 336. The same two SNPs are identified with all other methods except screen and clean. There is large agreement between the methods on which SNPs to include, although some methods detect more SNPs than others. Screen and clean is the exception: it identifies SNPs that are not found with other methods, not even the Fisher test which identifies by far the most SNPs (24). Only four out of the nine SNPs found with screen and clean are identified with any of the other methods.

\section{Discussion} \label{sec:discuss}
This paper presents a thorough simulation study of the performance of various methods at recovering true sparsity patterns, specifically with application in genetic studies.  Many of these methods are not designed for the objective of true sparsity recovery but still performed well in recovering true predictors. For lasso, elastic net and the Fisher test, this is at the expense of a large number of false finds; stability selection and DET achieve the true finds without too many false finds. 
Once the correlation becomes high, specialised methods such as DET are needed. The main drawback of DET, however, is that performance does not improve notably with sample size.  In our simulations screen and clean does benefits most from larger sample sizes.

Stability selection performed extremely well.  At first glance, it seems that because of the bootstrapping approach used, even moderate correlation may cause problems for the method.  Whilst it does fail for very large correlations, datasets with neighbouring correlations as high as $\rho =0.95$ are handled well.  Indeed, until the correlation reaches this high level, stability selection also controls the type one error.  \cite{meinshausen08b} proved type one error control for exchangeable coefficients, the work here represents further empirical evidence that similar control is achieved in many non-exchangeable situations. Indeed, when the definition of a true find is relaxed to declaration of a predictor with correlation of at least 0.9 with the true predictor, stability selection performs best in all scenarios. 

As the density of information collected increases, measured predictors will become increasingly correlated.  So methods that are specifically designed to handle this, such as direct effect testing, will become more important. DET, however, only deals with binary predictors and response and therefore, novel methods that can handle highly correlated continuous variables are needed.

\bibliographystyle{apalike}
\bibliography{biblio-full}

\end{document}